\documentclass{svjour3}                     % onecolumn (standard format)
\usepackage{graphicx}
\begin{document}

\title{On mass polarization effect in three-body systems%\thanks{Grants or other notes
}


\author{I. Filikhin  \and R. Ya. Kezerashvili \and V.~M.~Suslov \and B. Vlahovic
       }


\institute{I. Filikhin, V. M. Suslov and B. Vlahovic \at
              North Carolina Central University, Durham, NC 27707, USA \\           
              \email{ifilikhin@nccu.edu}           %  \\
           \and
           R. Ya. Kezerashvili \at
             Physics Department, New
York City College of Technology, The
City University of New York, Brooklyn, NY 11201, USA \at
The Graduate School and University Center, The
City University of New York,
New York, NY 10016, USA
}

\date{Received: date / Accepted: date}

\maketitle

\begin{abstract}
We evaluate the mass polarization term 
of the kinetic-energy operator 
for different three-body nuclear $AAB$ systems by employing the method of
Faddeev equations in configuration space. For a three-boson system this term
is determined by the difference of the doubled binding energy of the $AB$
subsystem $2E_{2}$ and  the three-body binding energy $E_{3}(V_{AA}=0)$
when the interaction between the identical particles is omitted. In this case: 
 $\left\vert E_{3}(V_{AA}=0)\right\vert >2\left\vert E_{2}\right\vert$.
 In the case of a system complicated by isospins(spins), such as the kaonic clusters $%
K^{-}K^{-}p$ and $ppK^{-}$, the similar evaluation is impossible. For these
systems it is found that $\left\vert E_{3}(V_{AA}=0)\right\vert <2\left\vert
E_{2}\right\vert$.  A model with an $AB$ potential
averaged over spin(isospin) variables transforms the later case to the first one.
The mass polarization effect calculated within this model is essential for the
kaonic clusters. Besides we have obtained the relation  $|E_3|\le |2E_2|$ for
 the binding energy of the kaonic clusters.
% \PACS{21.85.+d \and 21.45.-v \and 11.80.Jy \and 13.75.Jz}

\end{abstract}

\section{Introduction}
\label{intro}
The mass polarization effect of the kinetic-energy operator is well known in
atomic physics \cite{Fischer,Bransden}. The kinetic energy operator in the
Schr\"{o}dinger equation for an $N$-electron atomic system with a finite
nuclear mass $M$ in the centre-of-mass coordinate system is comprised of two
parts: the kinetic energy term related to the introduction of the reduced
mass and the mass polarization term (MPT) $-\frac{\hbar ^{2}}{M}
\sum\limits_{i<j}\nabla _{i}\cdot \nabla _{j}$, where the indices $i$ and $j$
denote the $i$th and $j$th electrons, respectively. This term leads to the
shift of atomic spectra and Hughes and Eckart \cite{1930Hughes} in 1930 were
the ones who studied this effect. For decades, the mass polarization term
has been treated differently in calculations: in the approximation of an
infinitely heavy nucleus and using perturbation theory. In the approximation
of an infinitely heavy nucleus the contribution of this term is zero, while
the evaluation of this term within perturbation theory proved to be
unreliable, as is pointed out in \cite{Prasad1966}. However, the contribution
of the mass polarization term in atomic physics is always considered as a
small correction due to large mass of the core nucleus \cite{Y1999,NN2008}.
For description of charged excitons and biexcitons in condensed matter physics the contribution of the mass polarization term cannot be ignored
due to the comparable masses of electrons and holes and requires its careful
consideration \cite{BHR,ZKV,Co}.
In nuclear few-body physics such correction can be essential. In particular,
the mass polarization term of the three-body kinetic-energy operator can
play an important role in the study of nuclear interactions in double
hypernuclei \cite{FG2002,H2002}
like the $^6_{\Lambda\Lambda}$He considered within the three-body cluster model $\Lambda\Lambda\alpha$. 
If we write the Schr\"{o}dinger equation for a three-body $AAB$ system using
 the non Jacobian coordinate set 
 and neglect the MPT interaction
between two identical particles, we obtain the trivial solution that
binding energy is $2E_{2}$, where $E_{2}$ is the two-body $AB$ energy. 
The consideration of the MPT shifts the energy by adding the 
mass polarization energy. 
For a three-boson $AAB$ system, this contribution  
 can be evaluated as \cite{H2002}
\begin{equation}
\delta B=2E_{2}-E_{3}(V_{AA}=0),  \label{RK1}
\end{equation}%
where $E_{3}(V_{AA}=0)$ is the three-body energy of the $AAB$ system when
interaction between two identical particles is omitted. 
Note that the contribution (\ref{RK1}) is small for the $\Lambda\Lambda\alpha$ system due to the $B$-particle  mass factor dependence which is expressed as $m_A/m_B$, where $m_A$ and $m_B$ are masses of non identical particles, and $m_B>m_A$. This mass ratio is approximately equal to $1/4$  for the $\Lambda\Lambda\alpha$ system. When $m_B>>m_A$, the contribution of the term can be neglected  \cite{H2002}. 

Consideration of the mass polarization term is very important for the three-body $AAB$ system when mass ratio for non identical particles is not small, for example, for the kaonic cluster ${\bar K}{\bar K}N$ the mass ratio of the kaon and nucleon is about 1/2. However, there are examples \cite{D2015,DIM2015} in the literature when this term is ignored
within a theoretical analysis of the kaonic clusters by  proposing  that $E_{2}=E_{3}(V_{AA}=0)/2$. 

In the presented work we focus on  different nuclear  $AAB$ systems involving two identical and one distinguishable particle  to evaluate
the mass polarization term of the kinetic-energy operator. We distinguish bosonic-like systems from systems having
isospins(spins) dependent interactions.
  In the case of a system complicated by isospins(spins), such as the kaonic clusters $%
K^{-}K^{-}p$ and $ppK^{-}$, the  evaluation (\ref{RK1}) is impossible. For these
systems it is found  that $\left\vert E_{3}(V_{AA}=0)\right\vert <2\left\vert
E_{2}\right\vert$, which gives a handy lower bound of 2$E_2$ to the $E_3$ (see also \cite{BMRW} in this regard).
For this case, the approach with average $AB$ potential may be
applied to reduce it to a bosonic-like system and the mass polarization can be roughly evaluated by using Eq. (\ref{RK1}). 
  Our treatment is based on the Faddeev equations in configuration space. These equations allow us to
separate components of the total wave function corresponding to the different particle rearrangements
 and to show the effects related to the
 exchange of identical particles and the  difference of particle masses. The latter facts are hidden in each Faddeev
component that corresponds to the interaction of any two particles in the presence of
the third.

The paper is organized in the following way. In Sec 2 we present the
formalism of the Faddeev equations in configuration space for a three-body
system with two identical particles. We consider two cases, when the
identical particles are fermions or bosons. The Faddeev equations are
written for the cases of two identical bosons and two identical fermions in
the $s$-wave approach and we consider the corresponding spin-isospin
configurations, as well as an average potential approach. 
The analysis of the mass polarization energy for a three-boson system within the $s$-wave Faddeev approach
is given in Sec. 3. The explanation for the mass polarization term and mass polarization effect is presented in Sec. 4
based on the Schr\"{o}dinger equation for bosonic-like systems.
The results of
numerical evaluations for  the double ${\Lambda}$-hypernucleus $^{~~6}_{\Lambda%
\Lambda}$He, the kaonic clusters ${K^-}{K^-}p$ and $pp{K^-}$, and
nucleus $^{3}$H are presented and discussed in Sec. 5. The conclusions follow in
Sec. 6.

\section{Formalism}
\label{sec:1}

\subsection{Faddeev equations in configuration space}
\label{subsec1}

The wave function of the three--body system can be obtained by solving the
Schr\"{o}dinger equation. Alternatively, in the Faddeev method the total wave function is
decomposed into three components: $\Psi =\Phi _{1}+\Phi _{2}+\Phi _{3}$ 
\cite{FaddeevConfigurSpace,NF68,K86}~. The Faddeev components $\Phi _{i}$ correspond to
the separation of particles into configurations $i+(kl)$, $i\neq k\neq
l=1,2,3$. Each Faddeev component $\Phi _{i}=\Phi _{i}(\mathbf{x}_{i},\mathbf{y}_{i})$ depends 
on its own set of the Jacobi coordinates $\mathbf{x}_{i}$ and $\mathbf{y}_{i}$. The
components satisfy the Faddeev equations in the coordinate representation
written in the form: 
$$
\left( H_{0}^{i}+v_{i}(\mathbf{x}_{i})-E\right) \Phi _{i}(\mathbf{%
x}_{i},\mathbf{y}_{i})=
$$
\begin{equation}
=-v_{i}(\mathbf{x_{i}})\left( \Phi _{k}(\mathbf{x}_{k},\mathbf{y}_{k}%
)+\Phi _{l}(\mathbf{x}_{l},\mathbf{y}_{l})\right) ,\ \ i\neq k\neq l=1,2,3,
\label{fadnoy}
\end{equation}
where $H_{0}^{i}=-(\Delta _{\mathbf{x}_{i}}+\Delta _{
\mathbf{y}_{i}})$ is the kinetic energy operator and $v_{i}$ is the
potential acting between the particles $(kl)$, $i\neq k\neq l$. We refer to  (\ref{fadnoy})  as the differential Faddeev equations (DFE). 
The mass scaled Jacobi coordinates $\mathbf{x}_{i}$ and $\mathbf{y}_{i}$ are expressed in terms of the particle coordinates $\mathbf{r}_i$
and masses  $m_i$ as:

\begin{equation}
\begin{array}{c}
\mathbf{x}_{i}=\sqrt{\frac{2m_km_l}{m_k+m_l}}(\mathbf{r}_k-\mathbf{r}_l), \qquad
\mathbf{y}_{i}=\sqrt{\frac{2m_i(m_k+m_l)}{M}}(\frac{m_k\mathbf{r}_k+m_l\mathbf{r}_l}{m_k+m_l}-\mathbf{r}_i),\\ M=m_i+m_k+m_l.\\
\end{array}%
\label{Jc}
\end{equation}

The
orthogonal transformation between three different sets of the Jacobi
coordinates has the form: 
\begin{equation}
\left( 
\begin{array}{c}
\mathbf{x}_{i} \\ 
\mathbf{y}_{i}%
\end{array}%
\right) =\left( 
\begin{array}{cc}
C_{ik} & S_{ik} \\ 
-S_{ik} & C_{ik}%
\end{array}%
\right) \left( 
\begin{array}{c}
\mathbf{x}_{k} \\ 
\mathbf{y}_{k}%
\end{array}%
\right) ,\ \ C_{ik}^{2}+S_{ik}^{2}=1,  
\end{equation}
where 
$$
C_{ik}=-\sqrt{\frac{m_{i}m_{k}}{(M-m_{i})(M-m_{k})}}, \quad
S_{ik}=(-1)^{k-i}\mathrm{sign}(k-i)\sqrt{1-C_{ik}^{2}}.
$$

\subsection{Faddeev equations for $AAB$ system}
\label{subsec2}

\begin{figure}[ht]
\begin{center}
\includegraphics[width=14pc]{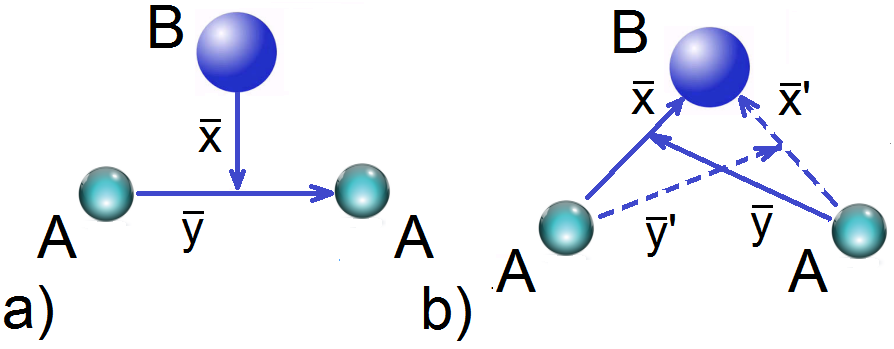}
\caption{\label{fig1}  
Schematic presentation of the $AAB$ system and rearrangements of Jacobi coordinates for the configurations: a) $(AA)B$, b) $A(AB)$, respectively.
}
\end{center}
\end{figure}

The objective of this work is a consideration of a three-body $AAB$ system
with two identical particles. Particularly, we focus on the kaonic clusters $%
ppK^{-}$\ and $K^{-}K^{-}p, $ $^{3}$H nucleus and  the double ${\Lambda }$%
-hypernucleus $_{\Lambda \Lambda }^{~~6}$He  in the framework of the $%
\Lambda \Lambda \alpha $ cluster model. Therefore, let us rewrite
 the system (\ref{fadnoy}) for a case of two identical particles. In this case the total wave function of the 
system is decomposed into the sum of the Faddeev components $U$ and $W$ corresponding to the $(AA)B$ and $A(AB)$ types of rearrangements: $\Psi =U+W\pm PW$, where $P$ is the permutation operator for two identical
particles. These types of the particle rearrangements  and corresponding Jacobi coordinates are graphically presented in Fig. \ref{fig1}.
 In the latter expression for $\Psi$, the sign ''$+ $'' corresponds to two identical bosons, while the sign ''$- $'' corresponds to two identical fermions, respectively. For a three--body system with two identical particles the
set of the Faddeev equations (\ref{fadnoy}) is reduced to the system of
two equations for the components $U$ and $W$ \cite{14,F2004}~: 
\begin{equation}
\begin{array}{l}
{(H_{0}^{U}+V_{AA}-E)U=-V_{AA}(W\pm PW),} \\ 
{(H_{0}^{W}+V_{AB}-E)W=-V_{AB}(U \pm PW),}
\end{array}
\label{GrindEQ__1_}
\end{equation}%
where the signs ''$+ $'' and ''$- $'' correspond to two identical bosons and fermions, respectively. The wave function of the system $AAB$  is symmetrized with respect to
 two identical bosons, while it is antisymmetrized with respect to two identical fermions. The partial wave analysis
of the DFE (\ref{GrindEQ__1_})  can be performed by
 the   $LS$ coupling scheme given in  \cite{K86,14,15}. 
The $LS$ basis allows us to restrict the model space to the states with the
total angular momentum $L=0$. 
%(when the spin projections of fermions are anti-parallel)

%and total isospin $I=\frac{1}{2}$. The possible isospin configurations $(AA)+B$ with $I(NN) = 0$ $(S(NN)=1) $ are not taken into account in our calculations. According to the evaluations of different authors the total contribution of the configuration is about 5\%~\cite{AYKN}. 

\subsection{Separation of spin(isospin) variables}
\label{subsec3}

The description of the aforementioned  $AAB$ systems differs for the type of ${AA}$ and ${AB}$ interactions. Without losing any generality and for the simplicity of the presentation we employ the $s$-wave spin-isospin dependent $V_{AA}$ and $V_{AB}$ potentials. This requires to write the DFE in the $s$-wave approach and consider the corresponding spin-isospin configurations. The separation of spin(isospin) variables leads to the Faddeev equations for the three-body system $AAB$ in the following form: 
\begin{equation} 
\label{eq:1} 
\begin{array}{l} 
   (H_0^U+V_{AA}-E)U
  =-V_{AA}D(1+p){\cal W}, \\
   (H_0^W+V_{AB}-E){\cal W}=-V_{AB} 
  (D^TU+Gp{\cal W}), 
\end{array} 
\end{equation} 
where matrices $D$ and $G$ are defined by the nuclear system under consideration, the ${\cal W}$ is a column matrix with the singlet and triplet parts of the $W$ component of the wave function of a nuclear system, and the 
exchange operator $p$ acts on the particles' coordinates only. 

Let us mentioned that the consideration of the spin and isospin dependence is
relevant for the \textit{AB} potentials in the neutron-proton and
proton-kaon cases for the $^{3}$H nucleus and kaonic clusters, respectively.
In the first case the potential and components of $\cal W$ are
labeled according to the pair spin. In the latter case the potential is isospin
dependent, but both channels have total spin $1/2$. Furthermore, spin/isospin
dependence is irrelevant to the \textit{AA} potential which is
assumed as  \textit{s}-wave interaction that is a spin-singlet $nn$ or $pp$ potentials or an isospin-triplet ${\bar K}{\bar K}$ potential.

For the $^3$H nucleus, considered as the $nnp$ system, 
the inputs into  (\ref{eq:1}) are the following: the spin singlet $nn$ 
potential $V_{AA}=v^s_{nn}$ and $V_{AB}=diag\{v^s_{np},v^t_{np}\}$ that
is a diagonal $2\times2$ matrix with the spin singlet $v^s_{np}$ and triplet $v^t_{np}$ 
 $np$ potentials, respectively, 
and 
\begin{equation}
\label{w1}
D=(-\frac12,\frac{\sqrt3}2),   \quad
G=\left( \begin{array}{rr}
-\frac12& -\frac{\sqrt3}2 \\
-\frac{\sqrt3}2& \frac12 \\
\end{array} \right),  \quad
{\cal W}=\left( \begin{array}{rr}
W^{s} \\ 
W^{t} \\ 
\end{array} \right),
\end{equation}
%We restrict the model space to the $s$-wave states with total spin $S=\frac12$.  
where $W^{s}$ and $W^{t}$ are the spin singlet and spin triplet components of the ${\cal W}$. For  a neutron-proton interaction, we use the semi-realistic Malfliet
and Tjon MT I-III \cite{MT} potential with the correction \cite{MT1}.  It has to be noted that we do  not use isospin formalism for the $nnp$ system. Thus, the protons and neutrons are not identical. The details of such treatment are presented in \cite{FSV16}.
%The potential has  two  components corresponding to  singlet and triplet spin states of the  pair $np$. 
%Below, we take $v^s_{nn}$=0.

\begin{figure}[h]
\centering
\includegraphics[width=9pc]{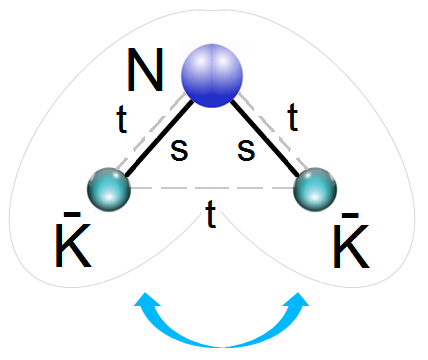}
\caption{\label{fig3} The isospin configurations in ${\bar K}{\bar K}N$  system. The isospin configurations for $NN{\bar K}$  system can be obtained by replacing
 $N\longleftrightarrow  {\bar K}$.
The pair  potential ${\bar K}N$ has singlet and triplet isospin component ($I=0$ or $I=1$). The ${\bar K}{\bar K}$ and $NN$ potentials are presented by triplet components. The exchange of identical particles shown by arrows. The  ${\bar K}N$ pair has a deep bound isospin singlet state.
}
\end{figure}

For $K^{-}K^{-}p$ and $ppK^{-}$,  despite of the fact that there are two identical bosons and two identical fermions, respectively, due to  symmetry of the spin-isospin configurations in  the
kaonic clusters, the $D$ and $G$ matrices in  (\ref{eq:1}) for these clusters are the same and have the following form  \cite{K2015}:
\begin{equation}
\label{w2}
D =(-\frac{\sqrt{3}}{2},-\frac{1}{2}), \quad G=\left( 
\begin{array}{cc}
\frac{1}{2} & \frac{\sqrt{3}}{2} \\ 
\frac{\sqrt{3}}{2} & -\frac{1}{2}%
\end{array}%
\right), \quad
{\cal W}=\left( \begin{array}{rr}
W^{s} \\ 
W^{t} \\ 
\end{array} \right).
\end{equation}
 Unlike to  (\ref{w1}) the  corresponding superscripts $s$ and $t$ in  (\ref{w2})  denote the isospin singlet $W^s$ and isospin triplet $W^t$ components of the ${\cal W}$, respectively.
For the $K^{-}K^{-}p$ kaonic cluster,  $V_{AA}=v^t_{{\bar K}{\bar K}}$ is the ${\bar K}{\bar K}$ potential in the triplet isospin state.
For the $ppK^{-}$ cluster ,  $V_{AA}=v^s_{NN}$ is the $NN$ potential in the singlet spin state. For both systems, one has to take $V_{AB}=diag\{v^s_{{\bar K}N},v^t_{{\bar K}N}\}$.
In the presented work, we used the $s$-wave Akaishi-Yamazaki
 (AY) \cite{YA07} and Hyodo-Weise (HW) \cite{HW,JK}  effective potentials for $\bar{K} \bar{K}$ and  $\bar{K} N$ interactions  
 which include the coupled-channel dynamics into a single
channel $\bar{K} N$ interaction. The graphical representation of the isospin configurations in ${\bar K}{\bar K}N$ and  $NN{\bar K}$ systems is given in Fig. \ref{fig3}. 

 The double-${\Lambda}$ hypernucleus $_{\Lambda\Lambda}^6$He is treated within the potential three-body cluster  $\alpha\Lambda\Lambda$ model using the frozen core approximation and thus, effects of core excitations are lacking. 
%The system is well clustering and could be treated as the $\alpha\Lambda\Lambda$ system.
%For the $_{\Lambda\Lambda}^6$He hypernucleus, considered in the cluster model as the three-particle $\alpha\Lambda\Lambda$ system 
For this case, one has 
\begin{equation}\label{w3}
D=G=1, \qquad {\cal W}=W 
\end{equation}
 in (\ref{eq:1}) and latter could be reduced to a scalar form \cite{F2004}.
For the  $\alpha\Lambda\Lambda$ calculations, we use modified Tang-Henrdon (TH(M)) potential  from   \cite{F2013} for the $\Lambda\alpha$ interaction. 
%as $\Lambda\alpha$ interaction.

\subsection{Average potentials}
\label{subsec4}

In this section, we consider the case when the ${AB}$ potentials  are spin(isospin) dependent. For example, 
in the $NN{\bar K}$ system, the ${\bar K}$ meson combines two nucleons into the bound state for
two ${\bar K}N$  isospin configurations which are energetically favorable. The effective 
${\bar K}N$ interactions have a strong attraction in the singlet $I=0$  channel 
and a weak attraction in the triplet $I=1$ channel. 
The ${\bar K}N$ pair is   bound in the singlet  state with the energy corresponding to 
one of the $\Lambda(1405)$ resonance. There is no a bound state in the triplet isospin state.

Below following  \cite{15,6}, we consider the effective potential obtained by averaging of the
initial potential over the spin variables, in the case of the $^{3}$H nucleus,
and the isospin variables, in case of
kaonic clusters.
 The isospin averaged potential $V^{av}_{{\bar K}N}$ is defined as: 
\begin{equation}
V^{av}_{{\bar K}N}=\frac{3}{4}v_{{\bar K}N}^{I=0}+\frac{1}{4}%
v_{{\bar K}N}^{I=1}.  
\label{av_pot}
\end{equation}
This potential has a moderate attraction in comparison with the strong attraction in the $
{I=0}$ channel. Note that this
simplification changes the two-body threshold,
which is not related to the ${K^-}p$ bound state as $\Lambda $(1405). 

Using the isospin (spin) averaging, Eqs. (\ref{eq:1}) can be reduced to the scalar form by an algebraic transformation.
Taking into account that $W=D{\cal W}$, $V_{AB}^{av}=DV_{AB}D^T$ and $DD^T=1$, $DV_{AB}GD^T=V^{av}_{AB}$ we obtain
\begin{equation} 
\label{eq:1av} 
\begin{array}{l} 
   (H_0^U+V_{AA}-E)U
  =-V_{AA}(1+p){W} , \\
   (H_0^W+V^{av}_{AB}-E){W}=-V^{av}_{AB} 
  (U+p{W}). 
\end{array} 
\end{equation} 
In this case, one can evaluate  the mass polarization term (\ref{RK1}) as
\begin{equation} \label{eq:2}
\delta B^{av}=2E^{av}_2-E^{av}_3(V_{AA}=0).
\end{equation}
 Here, $E_2^{av}$ is the two-body energy for the $AB$ pair with the averaged potential and $E^{av}_3(V_{AA}=0) $
 %$E_3^{av}$ 
 is the three-body energy with the averaged potential when the ${AA}$ interaction is omitted.
% that is noted as $E^{av}_3(V_{AA}=0)$. 
 The   three-body    $E_3^{av}(V_{AA}=0)$ energies are calculated by using  Eq.  (\ref{eq:1av}) with the averaged potential. 
  It has to be noted that  the two-body energies $E_2^{av}$ and $E_2$ are different due to the difference of the averaged and singlet ${\bar K}N$ potentials.  

  The averaged potentials for the $nnp$ (averaged in the spin space)  is constructed by the same way as in \cite{15}, while for the   ${\bar K}{\bar K}N$ systems (averaged in the isospin space) it is defined by  (\ref{av_pot}).
  
\section{When $E_3=2E_2$}
\label{sec2}

Let us consider the $s$-wave approach  for the Faddeev equations (\ref{eq:1av}) for the  $AAB$ system when 
 particles $A$ and $B$ are interacting with the  $V_{AB}$ potential. We assume that the interaction between two identical particles is omitted, 
therefore the potential 
%$AA$ potential is equal zero, 
$V_{AA}=0$. 
The $s$-wave DFE is reduced  toa single equation for the Faddeev component $W(x,y)$: 
\begin{equation} \label{DFE1}
(-\frac{\hbar ^{2}}{2\nu}
\partial _{y}^{2}-\frac{\hbar ^{2}}{2\mu}\partial _{x}^{2}+V_{AB}(x)-E)W(x,y)= 
-V_{AB}(x)
\int\limits_{-1}^{1}\frac12 du\frac{xy}{x^{\prime}y^{\prime }}W(x^{\prime},y^{\prime}),
\end{equation}
where $x$, $y$ are the Jacobi coordinates and $u$=$cos(\widehat{\mathbf{xy}})$, where $\widehat{{\mathbf{xy}}}$ is the angle between $\mathbf{x}$ and $\mathbf{y}$.
The identical particles in the system are labelled as 2 and 3 and $m_{2}=m_{3}=m$, while the $m_{1}$
is the mass of the $B$ particle.  
The appropriate transformation of coordinates and reduced masses are given by the following expressions: 
\begin{equation} \label{supplement11}
\begin{array}{l}
x^{\prime}=\left( (\frac{\mu}{m_{1}}x)^{2}+y^{2}-2\frac{
\mu}{m_{1}}xyu\right) ^{1/2},\quad \\
y^{\prime}=\frac{\mu}{
m_{1}}\left( (\frac{m_1}{\nu }x)^{2}+y^{2}+2\frac{m_{1}}{\nu}
xyu\right) ^{1/2}, 
\end{array}
\end{equation}
$$
\nu=\frac{m(m_{1}+m)}{m_{1}+2m}, \quad 
 \mu=\frac{m_{1}m}{m_{1}+m}.
$$ 
Two types of configurations for the particles in the $AAB$
system and Jacobi coordinates were  presented in Fig. \ref{fig1}. 
The configuration shown in Fig. \ref{fig1}b corresponds to one described by Eq. (\ref{DFE1}).

The analysis  of expressions (\ref{supplement11}) shows that when $m_1 >> m$, one can write  $x^\prime \approx y$, $y^\prime \approx x$ and $\nu \approx \mu$.
Within this approximation, after the integration by the variable $u$, Eq. (\ref{DFE1}) can be rewritten in the following symmetric form:
$$
(-\frac{\hbar ^{2}}{2\mu}
\partial _{y}^{2}-\frac{\hbar ^{2}}{2\mu}\partial _{x}^{2}+V_{AB}(x)+V_{AB}(y)-E)W(x,y)= 
$$ 
\begin{equation} 
=V_{AB}(y)W(x,y)-V_{AB}(x)W(y,x).
 \label{DFE2}
\end{equation}

By averaging both sides of (\ref{DFE2}) one obtains on the right hand side the integral  \\ $\int dxdy (W(x,y)V_{AB}(y)W(x,y)-W(x,y)V_{AB}(x)W(y,x))$. 
If the function $W(x,y)$ yields factorization and  $W(x,y)=W(y,x)$, then variables are separated and the values of this integral 
is equal to zero, due to the symmetry related to the replacement of variables $x\to y$, $y\to x$. 
In
this case, the function $W(x, y)$ is factorized as $W(x, y) = \phi(x)\phi(y)$. Taking into account that 
$
(-\frac{\hbar ^{2}}{2\mu}\partial _{x}^{2}+V_{AB}(x))\phi(x)~=~E_2\phi(x),
$
 one obtains the relation 
$
 E_3(V_{AA}=0)=2E_2 
$
 for the three-body ground state energy.  
 Let us note that the relation $  E_3^{av}(V_{AA}=0)=2E^{av}_2$ is valid for the spin(isospin) averaged potential as well.
 
One can separate two effects related to the mass polarization term. The first is the exchange 
effect which is related to the right hand side of (\ref{DFE2}) and includes the permutation operator of the identical particles
(see  (\ref{GrindEQ__1_})). In the limit of large $B$ particle mass the above mentioned integral is equal to zero.
In an opposite case, the latter  integral violates the relation $E_3=2E_2$. At the same time, the second effect is related to the  difference of the reduced masses $\mu$ and $\nu$.  
This difference violates $x$-$y$ symmetry on the left hand side of  (\ref{DFE2}).

\section{Mass polarization term and mass polarization  effect}

To better understand the  effect induced by the different masses of 
$A$ and $B$ particles in a three-body $AAB$ system one can use the non
Jacobian form of the Schr\"{o}dinger equation from \cite{H2002} that is
written in a self-explanatory notation to analyze the contribution of the
mass polarization term: 
\begin{equation}
\begin{array}{c}
(-\frac{\hbar ^{2}}{2\mu }\nabla _{r_{A_{1}}}^{2}-\frac{\hbar ^{2}}{2\mu }%
\nabla _{r_{A_{2}}}^{2}-\frac{\hbar ^{2}}{m_{B}}\nabla _{r_{A_{1}}}\nabla
_{r_{A_{2}}}+V_{AB}(r_{A_{1}})  \\ +V_{AB}(r_{A_{1}})-E)\Psi
(r_{A_{1}},r_{A_{2}})=0.  \label{Sh}
\end{array}
\end{equation}%
In the latter equation the third term is  the mass polarization term, $T_{MPT}=-\frac{\hbar^{2}}{m_B}\nabla_{r_{A_1}}\nabla_{r_{A_2}}$, the
interaction between two identical particles is omitted, $V_{AA}=0$, and  $%
E\equiv E_{3}(V_{AA}=0)$ corresponds to the binding energy of the $AAB$
system when the interaction between two identical particles is neglected.
The mass of each particle $m_A$, $m_B$ is always bigger than the reduced mass $\mu $ $%
m_{B}>m_A>\mu $\ and the reduced mass is always more close to the mass of
the lightest particle. In the case $m_A>m_{B}$ the contribution of the
mass polarization term can be the same order as the contribution of the
other two differential operators in Eq. (\ref{Sh}) due to the comparable
mass factors of these operators which are approximately $1/m_{B}$. In the
case $m_{B}>m_A$ the contribution of this term has the factor $1/m_{B}$,
while the mass factors of the other differential operators are about $1/m_A
$. When $m_{B}>>m_A$ the contribution of the mass polarization term can be
neglected.

 Within the first order of perturbation theory, when  $<T_{MPT}>$ $<<$ $|E_3(V_{AA}=0)|$ (case $m_B > m_A$) the initial  wave function is factorized as 
 $$\Psi(r_{A_1},r_{A_2})=\phi(r_{A_1})\phi(r_{A_2}),$$
 where $\phi(r_{A_1})$ ($\phi(r_{A_2})$) is a solution of  two-body Schr\"{o}dinger equation for the $AB$ subsystem.
 Averaging of Eq. (\ref{Sh}) leads to the following relation between $E_2$ and $E_3(V_{AA}=0)$:
\begin{equation} 
<T_{MPT}>=2E_2-E_3(V_{AA}=0).
 \label{mp}
\end{equation}
We have obtained the evaluation for the mass polarization term when $m_B > m_A$.
For the simplest case, when the MPT is ignored in Eq.  (\ref{Sh}) , $<T_{MPT}>=0$,  we have $E_3(V_{AA}=0, {\textrm {without MPT}})=2E_2.$
For the general case, the effect of the mass polarization term can be estimated as 
 $$E_3(V_{AA}=0, {\textrm {without MPT}})-E_3(V_{AA}=0).$$
This estimation  is valid for any mass ratio $m_B/m_A$. 
The relation (\ref{mp}) is known in nuclear physics as the mass polarization effect \cite{H2002} expressed as the deference of
$2E_2$ and $E_3(V_{AA}=0)$ according Eq. (\ref{RK1}). Therefore, the $\delta B$ in  Eq. (\ref{RK1})
is a  direct estimation of the MPT for bosonic-like systems, $\delta B=<T_{MPT}>$, when $m_B > m_A$.

Let us mention that the MPT is not an artefact of not using Jacobi coordinates. In the system of
reference presented in Eq. (\ref{Sh}) this is a kinematical effect related
to the presence of the third particle $A$ when the other one interact with
the particle $B$. The presence of the third particle gives the
redistribution of kinetic energy and as a result $AB$ subsystem is off the
energy shell. It is well know that a physical result does not depend on
the system of references. If one considers the $AAB$ using Jacobi
coordinates by employing the Faddeev equations the latter fact is hidden in
each Faddeev component that corresponds to the interaction of any two
particles in the presence of the third.  

\section{Numerical  Results}
\label{sec3}

\subsection{Bosonic-like system}
\label{subsec31}

Let us consider a three-boson system to exemplify the formalism presented above.
When $m_1 > m$, the value  $\delta B$ is mainly determined by the right hand side of  (\ref{DFE2}).  
To illustrate this statement, we {\normalsize %
consider the $_{\Lambda \Lambda }^{~~6}$He nucleus within the cluster model
as a three-body $\Lambda \Lambda \alpha $ system} and show a correlation
between a type of $AB$ potential and the mass polarization term $\delta B$. 
We assume  the frozen core approximation (there is no dynamical
change of the core-nuclear structure) and the $s$-wave approach is based on Eq. (\ref{eq:1}) with
the definitions  (\ref{w3}). We note this model as ''bosonic-like''  due to a similarity of these equations to ones for a system of three bosons. 
In Eq. (\ref{RK1}) ${\normalsize E}_{{\normalsize 2}}
${\normalsize \ is the ground state energy of the $_{\Lambda }^{5}\mathrm{He}
$ nucleus within the two-body cluster model}. $E_{3}(V_{\Lambda \Lambda
}=0)$ is {\normalsize the ground state energy of $_{\Lambda \Lambda }^{~~6}%
\mathrm{He}$ nucleus in the framework of the three-body cluster model $%
\Lambda \Lambda \alpha $ when $V_{\Lambda \Lambda }~=~0.$ In previous
calculations \cite{FG2002,H2002} it was determined that }${\normalsize 2E}_{%
{\normalsize 2}}-E_{3}(V_{\Lambda \Lambda }=0)\neq 0$ and {\normalsize the
latter is related to the effect of the mass polarization term of the
kinetic-energy operator $\delta B_{\Lambda \Lambda }$.
Results of our calculation and calculations \cite{F2004} for the $\Lambda\Lambda\alpha $ 
system are presented in Table \ref{t0}.  
In  our previous calculations \cite{F2004} we consider the several $\Lambda\alpha$ potentials. These potentials 
have different shapes, while reproduce  closely the experimental value of the binding energy for the $_{\Lambda}^5$He hypernucleus. 
%The $\Lambda\alpha$ potentials have different types \cite{F2004}. 
The difference of the $\Lambda\alpha$  potentials can be clarified by indicating 
 the corresponding $\Lambda\alpha$ scattering lengths. In particular, the  scattering length  characterizes the behavior of the potential of pair interaction  at 
large distances. 
As example, in Table \ref{t0}, the Tang-Herndon (TH) potential \cite{TH} is an attractive potential with no  repulsive core. The Isle potential \cite{Isle} has a weak repulsive core and
 decreases slowly at large distances. 
In Table \ref{t0}, the potentials are arranged  by increasing the values of the scattering length  
from 3.63~fm for the TH potential  to 4.24~fm  for the potential Isle, consequently. 
As follows from Table \ref{t0} the similar pattern appears for  the  ${\delta B}/{-E_{3}}$ that is calculated for the same set of potentials. Thus, the  mass polarization energy has the essential 
dependence on the type of  $\Lambda\alpha$ interaction. 
However, there is no correlation  between $\delta B$ and  $E_2$ within the $\Lambda\Lambda\alpha $  cluster model 
as this follows from Table \ref{t0}.
\begin{table}[t]
\caption{\label{t0}  The two-body $E_2$ and three-body $E_3$ ground state energies of the $\Lambda\alpha$ ($_{\Lambda}^5$He) and  $\Lambda\Lambda\alpha $ ($_{\Lambda\Lambda}^6$He) systems 
 for different $\Lambda\alpha$ potentials. The $\Lambda\Lambda$ interaction is omitted in~(\ref{eq:1}), $V_{\Lambda\Lambda}$=0, and the
mass polarization term is evaluated as $\delta B=2E_2-E_3(V_{\Lambda\Lambda}=0)$. 
$a$ is the  $\Lambda\alpha$ scattering length.
}
\begin{tabular}{@{}ccccccc@{}} \hline\noalign{\smallskip}
$\alpha\Lambda$ potential: & TH\cite{TH} &  TH(M)\cite{F2013} & Gibson\cite{Gibson} & MS\cite{MS} & MSA\cite{MSA} & Isle\cite{Isle} \\ \noalign{\smallskip}\hline\noalign{\smallskip}
$E_{2}$, (MeV)&-3.03$^{*}$$^{\rm a}$ &-3.12 &-3.08$^{*}$ &-2.84$^{*}$&-3.12$^{*}$ &-3.10$^{*}$ \\
$a$, (fm) &  3.63$^{*}$& 3.70 & 3.80$^{*}$ & 4.00$^{*}$& 4.18$^{*}$ & 4.24$^{*}$\\
$E_{3}$, (MeV) & -6.335$^{*}$ & -6.49 & -6.383$^{*}$ &-5.890$^{*}$& -6.409$^{*}$&-6.341$^{*}$\\
$\delta B$, (MeV)& 0.275 & 0.25 &0.223 &0.210 & 0.169 & 0.141\\
$-{\delta B}/{E_{3}}$ & 4.3\% & 3.8\% & 3.5\% &  3.5\%&  2.6\%& 2.2\%\\
\noalign{\smallskip}\hline
\end{tabular}\\
$^{\rm a}$  The values marked by stars are taken from \cite{F2004}.
\end{table} 

\subsection{Kaonic systems}
\label{subsec23}

\subsubsection{ ${\bar K}{\bar K}N$ system\\}
The ground state energy $E_3$ of the ${\bar K}{\bar K}N$ system  
was calculated using the effective  AY and HW potentials describing the ${\bar K}N$  and $\bar K\bar K$ interactions. 
These effective ${\bar K}N$ $s$-wave potentials implicitly include $\pi\Sigma$  coupling and are widely used for description of the few-body kaonic clusters. The numerical results are presented in Table \ref{t1}. 
In Table \ref{t1} we compare our results with the results  obtained in  \cite{6} within a variational approach with a Gaussian
expansion method. The small discrepancies in calculations are related to the different    $K$-meson mass used in the calculations.
We used the value 493.677~MeV for the $K^-$ mass  from \cite{O14},  while in \cite{6} the authors used the mass 495.7~MeV.
The consideration of the same mass as in \cite{6} changes   $E_2$ energy for the AY potential from our  value of -30.3~MeV to  the value of~-30.6~MeV \cite{6}. 
However, $E_3$=-31.66~MeV does not become -32.3~MeV as in \cite{6}. Results of our calculation  are different from
ones reported in  \cite{6} by 0.3~MeV for both versions of the  ${\bar K}{\bar K}$ AY potentials: AY(104) and AY(70).
To check the accuracy of our calculations, we compare our results for the $nnp$ system and one reported in \cite{FGP}.
For $E_3$ energy we obtained the value of -8.534~MeV that is very close to the value of -8.535~MeV \cite{FGP},
when the MT I-III nucleon-nucleon potential is used.
Note that the computer codes for $E_3$ calculation are  the same for the both  $nnp$ and ${\bar K}{\bar K}N$ systems with
taking into consideration the exchange for potentials, masses,  matrices   $D$ and  $G$ in (\ref{eq:1}).  

One can see from Table \ref{t1} that the energy calculated  under the condition $V_{{\bar K}{\bar K}}=0$
has the larger absolute value comparing with one obtained within the full potential model that includes all interactions between particles. This is possible due to the repulsive  $K^- K^-$ interaction for the both
 AY and HW potentials. The absolute value of the  ground state energy $E_3$ is larger than one for the ground state energy  $E_2$ of the $K^- p$ pair (singlet isospin state)  for both ${\bar K}N$ potentials and for the both cases  $V_{\bar K\bar K}=0$ and $V_{\bar K\bar K}\ne0$ so that the relation $2E_2-E_3<0$ is satisfied.
\begin{table}
\caption{\label{t1} The ground state energies: $E_2$ for the $K^-p$, and $E_3$ for the ${\bar K}{\bar K}N$  system  with the AY and HW potentials for the $\bar{K} \bar{K}$ and  $\bar{K} N$ interactions.
The two-body  $E_2^{av}$ and three-body  $E_3^{av}$ energies are presented for the averaged potential $\frac{3}{4}v^s_{{\bar K}N}+\frac{1}{4}v^t_{{\bar K}N}$. The mass polarization term of the three-body kinetic-energy operator is evaluated by applying (\ref{eq:2}). $m_K$ is the $K^-$-meson mass used for the calculations. The averaged nucleon mass of 938.9 MeV is used as the input for the proton mass. All entries are given in MeV.
}
\begin{tabular}{llccccc}\\ \hline\noalign{\smallskip}
&$\bar{K}\bar{K}$,   $\bar{K}N$   & $E_{2}$ &$E_{3}$&$E_2^{av}$&$E_{3}^{av}$ & $\delta B^{av}$\\ \noalign{\smallskip}\hline\noalign{\smallskip}
$m_K$=493.667&AY , AY  & -30.3& -31.66$^{\rm a}$\cite{KTFSV}&-9.63 & -16.1 & --\\
&$V_{\bar K \bar K}=0$ , AY   & &-35.18  &  & -21.7 & 2.44\\
&HW , HW  &-11.16 & --$^{\rm b}$ &-1.11 & -1.62& --    \\
&$V_{\bar K \bar K}=0$ , HW   & &-12.18  & & -3.07 & 0.85\\
\noalign{\smallskip}\hline\noalign{\smallskip}
$m_K$=495.7 &   AY , AY    &-30.6  & -32.0 & & &\\
                     &  $V_{\bar K \bar K}=0$ , AY    & &  -35.6 & & &\\
                     &  HW , HW   &-11.42  &   --$^{\rm b}$& & &\\
                     &$V_{\bar K \bar K}=0$ , HW  &  &  -12.5 & & &\\
  \cite{6}&   AY , AY    &-30.6 & -32.3 & & &\\
&  $V_{\bar K \bar K}=0$ , AY      & &  -36.0 & & &\\
&  HW , HW   &-11.40 &  -11.4& & &\\
&$V_{\bar K \bar K}=0$ , HW  &  &  -12.6& & &\\
\noalign{\smallskip}\hline
\end{tabular} \\
$^{\rm a}$R. Ya. Kezerashvili, S. M. Tsiklauri, I. N. Filikhin, V. M. Suslov, and B. Vlahovic, {\it $ppK^-$ and $K^-K^-p$ Clusters}, 
\emph{The 21st International Conference on Few-Body Problems in Physics (FB21)}, Chicago, Illinois, USA, May 18-22, 2015. \\
$^{\rm b}$ no bound state.
\end{table} 

The model with averaged potential demonstrates the opposite relation 
between $E_3(V_{\bar K\bar K}=0)$ and $E_2$:  $2E_2^{av}-E_3^{av}(V_{\bar K\bar K}=0)>0$. 
The isospin components of the wave function  present in (\ref{eq:1})  with different coefficients due to the non-trivial matrix $G$ and $D$ (''isospin complication''). 
  Within the averaged potential approach, both kaons interact with  the proton by the same average potential.  There is a similar case when the singlet potential is equal to the triplet ${\bar K} N$ potential.  Eq. (\ref{eq:1}) is  reduced also to one with the trivial $G$ and $D$ matrices. 
  
\subsubsection{$NN{\bar K}$ system:\\}
In Table \ref{t1b}, we compare our results for the $NN{\bar K}$ system with the  variational calculations from \cite{DHW09} and  \cite{YA07}. The results are in acceptable agreement taking into account the difference of the models and methods. The details of  discrepancy for the ground state energies -39~MeV and -48~MeV presented in Table \ref{t1b} are discussed in  \cite{DHW09}.
\begin{table}[t]
\caption{
\label{t1b} Ground state energies $E_3$ of the  $NN{\bar K}$ system (in MeV) with the AY and HW potentials for the
${\bar K}N$ interaction and T (the Tamagaki potential)\cite{YA07} and simulating AV18 potential (sAV18)\cite{DHW09} for the singlet $NN$ interaction. 
}
{\begin{tabular}{llccc}  \hline\noalign{\smallskip}
  $NN$ & ${\bar K}N$  &          & Ref.\cite{DHW09}&  \cite{YA07} \\ \noalign{\smallskip}\hline\noalign{\smallskip}
          T & AY  &-46.35 &-39.0  & -48\\
   sAV18 & AY  & -46.21  & -45.8$^{\rm a}$ \cite{DIM2015} & --\\       
   sAV18 & HW & -20.57  & -20$\pm $3$^{\rm b}$  & --\\     
\noalign{\smallskip}\hline
\end{tabular} }\\
$^{\rm a}$  Energy dependent $\bar{K}N$ potential defined by the ansatz $M_N + m_K -B_K/2$.\\
$^{\rm b}$   Different chiral ${\bar K}N$  potentials.
\end{table} 

According to  (\ref{eq:1}), each spin(isospin) configuration of the system is represented by the corresponding Faddeev component of the wave function. 
The relative contributions of the isospin Faddeev components to the total wave function  
of the  $NN{\bar K}$ system can be seen from  Fig. \ref{fig2}.  The  model \cite{K2015} with the AY and MT I-III potentials is used for the calculations.
The first configuration  $N(N{\bar K})$ with the singlet isospin state of the pair (${\bar K}N$)
dominates with the maximum value of the $W^s$ component about 1.  
The area of rearrangement for the wave function of the $({\bar K}N)N$ singlet configuration, 
 shown in Fig. \ref{fig2}a,  is visibly restricted along the coordinate $x$ (the distance between ${\bar K}$ and $N$) due 
 to the existence of the strong bound ${\bar K}N$ state. At the same time, the function is quite prolongate along the $y$ axis due to the
 relatively weak bound of the third particle ($N$) with the pair ${\bar K}N$ having the energy  $|E_3-E_2|<|E_2|$. The numerical solution reported in  \cite{K2015}  gives  $E_3$=-46.01~MeV with $E_2=$-30.26~MeV.
\begin{figure}[h]
\centering
\includegraphics[width=29pc]{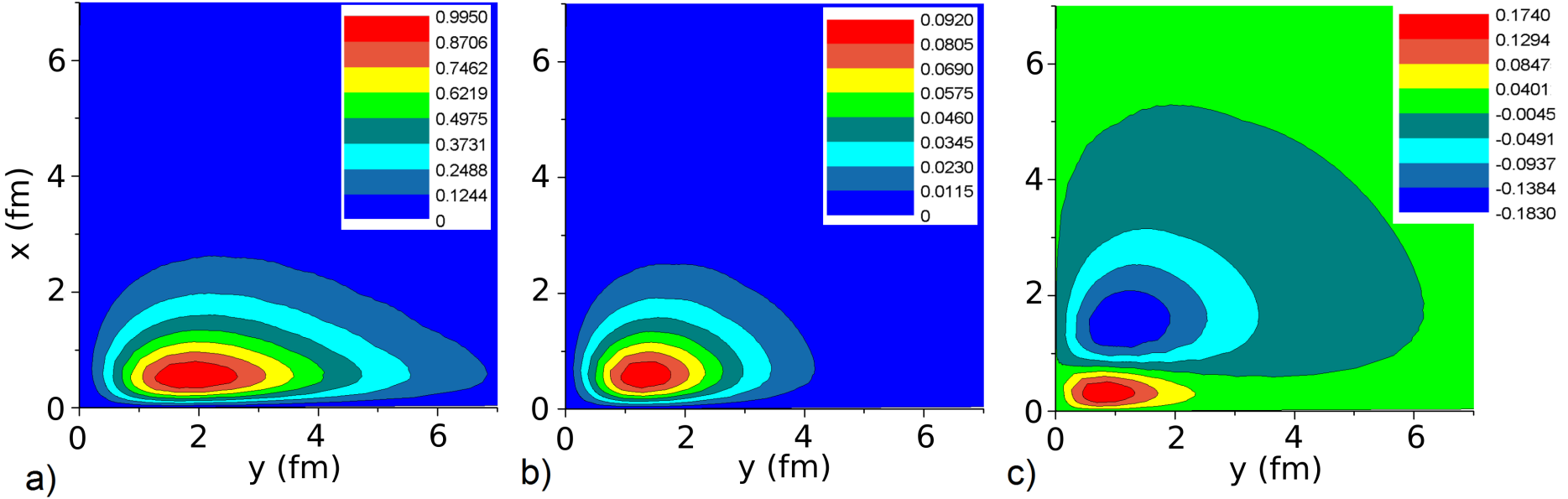}
\caption{\label{fig2}
Contour plots of the $s$-wave  Faddeev components in the kaonic $NN{\bar K}$ system for the configurations
of subsystems: a)  (${\bar K}N$)$N$ with the singlet isospin state of the pair (${\bar K}N$); b) (${\bar K}N$)$N$
with the triplet isospin state of the pair (${\bar K}N$); c) $($NN$){\bar K}$. 
}
\end{figure}

The second configuration
$({\bar K}N)N$, shown in Fig. \ref{fig2}b, with the triplet isospin state of the ${\bar K}N$ pair  is suppressed 
by the factor of 10 and the maximum of this component does not exceed  the value of 0.1. 
The maximum of the $W^t$ component of the $(NN){\bar K}$ configuration presented in Fig. \ref{fig2}c is relatively larger than the maximum of the component presented in Fig. \ref{fig2}b. 
   The similar picture was obtained for the $K^-K^-p$ wave function. The similarity is due to the use of the same Eq.  (\ref{DFE1})  for the description of the both  $K^-K^-p$ and  $pp{K^-}$ clusters. 

 The  small contribution of the  triplet isospin $W^t$ component for the $({{\bar K}N})N$ ($({{\bar K}N}){\bar K}$ ) configuration to the total wave function, as is illustrated in Fig. \ref{fig2}, motivated us to evaluate  $E_3$ by 
 neglecting the triplet component  of ${\bar K}N$ potential along with the nucleon-nucleon potential and, thus, 
 keeping the singlet  ${\bar K}N$ potential only as non-zero one. In the case of the $K^-K^-p$ cluster, the potentials of the triplet ${\bar K}N$ state and  ${\bar K}{\bar K}$ pair are zero.
 As a result, 
 the singlet $W^s$ component only  is  a non trivial in Eq. (\ref{DFE1}).  
%    The right hand side of the equation for  the component $W^s$ includes ''exchange'' term  related to the  permutation %operator.  This term provides an additional attraction and the system is bound. 
 % It may be described as an ''exchange effect''  existing in the three-body system with two identical particles. However,  %an analog of this ''exchange'' term is appeared in the Faddeev equations for any  three-body system.
%  The same situation takes place for the $NN{\bar K}$ system.
%This  ''exchange effect" is more visible when heavy particles are involved. 
  The results of calculations are presented in Table \ref{t1a}. Analysis of the results shows that  $K^-K^-p$ and $ppK^-$ clusters are 
  still bound by the singlet component of the ${\bar K}N$ potential when the triplet component of the  potential is equal to zero, and the ${\bar K}{\bar K}$($I=1$) and $NN$ 
 potentials are equal to zero, respectively for the corresponding  cluster. This is due to the domination of the singlet component
 of the ${\bar K}N$ potential over the triplet one. The results shown  in Table  \ref{t1a} support the relation $2E_2-E_3<0$  for the both cases:  $V_{NN}=0$ and $V_{NN}\ne 0$. 

For description of the $NN{\bar K}$ cluster  we take into account  the contribution of the $s$-wave of  $NN$ potential.  When the spin state of the two protons is restricted to $S = 0$, the orbital momentum of the $NN$ pair is $l = 0, 2, 4$. As is shown in Ref. \cite{K2015}, the contributions of the higher orbital are small enough and the $s$-wave consideration is  reasonable.   
         
 \subsubsection{The relation $|E_3|\le |2E_2|$ \\ }
 For the both  kaonic clusters, we found that $2E_2-E_3(V_{AA}=0)<0$ (see Table \ref{t1a}). This is due to the difference of the strengths of  the $ {\bar K}N$ potential components for the $I=0$ and $I=1$.
When $v^{s}_{{\bar K} N}=v^{t}_{{\bar K}N}=v_{AB}$, Eq. (\ref{eq:1}) reduced to the "scalar" form (\ref{eq:1av})  in the same way as using the $D$-matrix transformation and taking into account that $V_{AB}=v_{AB}I$, where $I$ is the identity matrix. In this case the definition of the averaged potential is not necessary and the 
relation $2E_2-E_3(V_{AA}=0)>0$ (the same as for the $\Lambda$$\Lambda\alpha$ case) will be automatically  satisfied.     
%If  the nucleon-nucleon  interaction is omitted  the corresponding $-E_3(V_{pp}=0)$ energy is 42.94~MeV that means %the first proton is bound with energy of 30.3 MeV and the second one is bound with the energy of 42.9-30.3=12.6~ MeV %as one can  see from Table  \ref{t1a}.  One has to extract the mass polarization energy from the latter value also. The %resulting energy picture  is far from the symmetrical ''particle picture'' presented in \cite{D2015,DIM2015}, where %protons are bound with the kaon with the same energies (42.9/2=21.45~MeV). 

We have to note here, that the value of $2E_2-E_3(V_{AA}=0)$  has exact physical interpretation for the spin/isospin averaged approach. The value estimates the mass polarization term of the kinetic operator by Eq. (\ref{eq:2}).
For general case of the spin/isospin dependent systems, this value is negative.
It is important to mention that for the binding energy of the
kaonic cluster $NN{\bar K}$  (${\bar K}{\bar K}N$) $|E_3|\le |2E_2|$  due to the weakly attractive (repulsive)  $AA$ potential.  In particularly,  $|E_3|$ somewhat is  increased relatively the value of  $|E_3(V_{NN}=0)|$ by the attracting $NN$ force and with account  of other possible physical channels \cite{D17}. 

\begin{table}[t]
\caption{
\label{t1a} Ground state energies $E_3$ of the ${\bar K}{\bar K}N$   and $NN{\bar K}$  clusters (in MeV) with the AY  potential for the
${\bar K}N$ interaction and the MT I-III potential for the $NN$ interaction. Results for the HW potential are given in parentheses. 
The difference of  the two-body  $2E_2$ and three-body  $E_3$ energies are presented.
}
{\begin{tabular}{llcc} \hline\noalign{\smallskip}
System&   Potentials     & $E_{3}$  &$2E_{2}-E_{3}$ \\ \noalign{\smallskip}\hline\noalign{\smallskip}
 ${\bar K}{\bar K}N$   & $V_{{\bar K}{\bar K}}=0$, $v_{{\bar K}{N}}^{t}=0$& -31.62 &-28.90 \\ 
                 & $V_{{\bar K}{\bar K}}=0$, AY& -35.18  &-25.34\\ \noalign{\smallskip}\hline
$NN{\bar K}$  & $v_{NN}=0$, $v_{{\bar K}{N}}^{t}=0$ &-36.15 (-12.46) & -24.37  (-9.86)\\
               & $v_{NN}=0$, AY &-42.94 (-17.11)  & -17.58 (-5.21)\\             
              &  MT I-III, $v_{{\bar K} {N}}^{t}=0$ &-41.47 (-17.08)  & -19.15 (-5.24)\\
              &  MT I-III, AY &-46.01\cite{K2015} (-20.46)  & -14.51 (-1.86)\\     
\noalign{\smallskip}\hline
\end{tabular} }
\end{table} 

\subsection{Comparison of mass polarization effect for several $AAB$ systems}
\label{subsec33}
To illustarate a correlation between the mass ratio $m_A/m_B$
and the contribution  of the mass polarization term to the three-body ground energy, 
we have calculated the ground state energy for different systems under the condition $V_{AA}=0$.
The considered systems  are ${K^-}{K^-}p$, $pp{K^-}$,
 $^3$H and $_{\Lambda\Lambda}^6$He.  
The models with the averaged potentials are used for the calculation. 
The mass ratios $m_A/m_B$ are rounded to fractions and the corresponding mass ratio  $m_A/m_B$ varies in the range from 1/4 to 2. 
In  Table \ref{t2} we  have presented the results of calculations for  the two-body $E_2^{av}$ and three-body $E_3^{av}$ ground state energies. The mass polarization term of the three-body kinetic-energy operator  is evaluated as $\delta B^{av}$ according to  Eq. (\ref{eq:2}).  
 When $m_A<m_B$ there is a strong correlation of the $-{\delta B^{av}}/{E_{3}^{av}}$ with $m_A/m_B$.
The value of   $-{\delta B^{av}}/{E_{3}^{av}}$ increases when the ratio $m_A/m_B$ increases from $\approx$ 1/4 to $\approx$ 1.  
%the $B$-particle mass  decreases to the limit $m_A/m_B$=1 and reaches  20\% of the  $E_3$ value.
The case when $m_A>m_B$ takes place for the $pp{K^-}$ cluster. The analysis of the results in Table \ref{t2} leads to the conclusion that the relative contribution of $\delta B^{av}$ into  $E_3$ energy is larger than one for the ${ K^-}{ K^-}p$ cluster.
\begin{table}[ht]
\caption{The two-body $E_2^{av}$ and three-body $E_3^{av}$ ground state energies (in MeV) are presented for different systems with averaged potentials (excluding the $\Lambda\Lambda\alpha $ system, as a system described by ''scalar'' equation).  The mass polarization term of the three-body kinetic-energy operator $\delta B^{av}$ is  evaluated in MeV using  (\ref{eq:2}).
The mass ratio $m_A/m_B$ is rounded to a fraction.
\label{t2}}

\begin{tabular}{@{}clcccc@{}} \hline\noalign{\smallskip}
System   & ${m_A}/{m_B}$ & $E_{2}^{av}$ & $E_{3}^{av}$ & $\delta B^{av}$ &$-{\delta B^{av}}/{E_{3}^{av}}$\\
\noalign{\smallskip}\hline\noalign{\smallskip}
$nnp$ ($V_{NN}=0$) & $\approx$ 1  & -1.53 & -3.87 & 0.81 & 21\%\\
${ K^-}{ K^-}p$ ($V_{\bar K \bar K}=0$)&  $\approx$ 1/2 & -9.63 & -21.7 & 2.44& 11\% \\
$\Lambda\Lambda\alpha $ ($V_{\Lambda\Lambda}=0$)&  $\approx$ 1/4& -3.12 & -6.49 & 0.25& 4.3\% \\   \hline
\noalign{\smallskip}
$pp{K^-}$ ($V_{NN}=0$) & $\approx$ 2/1  &-9.63 & -24.7 & 5.44 & 22\%   \\
\noalign{\smallskip}\hline
\end{tabular}
\end{table} 

\section{Conclusions}
In this paper, we have considered several  three-body $AAB$ systems   with two identical particles.
In the case of the systems described by the scalar form of Eq. (\ref{eq:1}), the mass ratio $m_A/m_B$ correlates clearly
with  the mass polarization term for the three-body energy. This contribution also weakly depends on the   $AB$ potential and correlates with  the $AB$  scattering length.  We have shown that the additional energy related to  the mass polarization term is exactly estimated using Eq. (\ref{RK1}) for any mass ratio $m_A/m_B$.

 The relation (\ref{RK1}) cannot be satisfied for a three-body system $AAB$ with a spin(isospin) dependent $AB$ interaction, such as the kaonic clusters ${\bar K}{\bar K}N$ and $NN{\bar K}$.
The  "isospin complication" leads to the following evaluation for three-body ground state energy of the kaonic clusters:
%\begin{equation}\label{E3}
$ |E_2|< |E_3(V_{AA}=0)|<|2E_2|$.
%\end{equation}
The relation  gives the upper value which can be reached by using  isospin formalism.
For the  $ppK^-$ cluster,  $|E_3|$ is
slightly larger then  $|E_3(V_{NN}=0)|$ due to the weakly attracting $NN$ force and taking into account  other possible physical channels. However,  $|E_3|$  is less than $|2E_2|$ and is essentially less than  the
experimentally motivated  value  about 100~MeV  \cite{I,G2016}. 

For kaonic  clusters, the configuration   with the singlet isospin state of the pair ${\bar K}N$
dominates. This makes possible that the $K^-K^-p$ and $ppK^-$ systems can be 
bound in the  cases when the pair ${\bar K}{\bar K}$($I=1$) and triplet ${\bar K}N$ potentials are equal to zero ($K^-K^-p$ cluster) and when the $NN$ and triplet ${\bar K}N$ potentials are equal to zero ($ppK^-$ cluster).

The mass polarization effect for the kaonic  clusters evaluated using the averaged potential approach  is essential. It has to be taken into account  when one attempts to construct a two body $\bar{K}N$ potential using a relation between two-  and three-body binding energies within the ''particle picture'' approach \cite{D2015}.

Finally it should be mentioned that we present the calculations for the kaonic system in the framework of the approximation with the effective complex $\bar{K}N$  potential. The complete treatment of the $\bar{K}NN$ system should be done within a coupled channel approach that  explicitly includes effects due to  the $\pi\Sigma$  coupling. Such consideration can lead to a possible modification of the results of our calculations. However, this will not change the qualitative conclusion that follows from our approach with the effective $\bar{K}N$ potential which implicitly includes the $\pi\Sigma$  coupling.

\begin{acknowledgements}
This work is supported by 
the National Science Foundation grant Supplement to the NSF grant HRD-1345219 and NASA (NNX09AV07A). R.Ya. K. partially supported by MES RK, the grant 3106/GF4.
\end{acknowledgements}

%\section*{References}

\end{document}